# iTRI-QA: a Toolset for Customized Question-Answer Dataset Generation Using Language Models for Enhanced Scientific Knowledge Retrieval and Preservation

Qiming Liu[1], Zhongzheng Niu[2], Siting Liu[3], Mao Tian [4*]

[1] Computer Science Department, University of California, Los Angeles.

[2] Department of Epidemiology and Environmental Health, University at Buffalo.

[3] Mathematics Department, University of California, Riverside.

[4] Jonsson Comprehensive Cancer Center, University of California, Los Angeles.

[*] Correspondence: maotian@mednet.ucla.edu

## Abstract

The exponential growth of AI in science necessitates efficient and scalable solutions for retrieving and preserving research information. Here, we present a tool for the development of a customized question-answer (QA) dataset, called Interactive Trained Research Innovator (iTRI) - QA, tailored for the needs of researchers leveraging language models (LMs) to retrieve scientific knowledge in a QA format. Our approach integrates curated QA datasets with a specialized research paper dataset to enhance responses' contextual relevance and accuracy using fine-tuned LM. The framework comprises four key steps: (1) the generation of high-quality and human-generated QA examples, (2) the creation of a structured research paper database, (3) the fine-tuning of LMs using domain-specific QA examples, and (4) the generation of QA dataset that align with user queries and the curated database. This pipeline provides a dynamic and domain-specific QA system that augments the utility of LMs in academic research that will be applied for future research LM deployment. We demonstrate the feasibility and scalability of our tool for streamlining knowledge retrieval in scientific contexts, paving the way for its integration into broader multi-disciplinary applications.

# Introduction

The rapid expansion of scientific literature has created a significant challenge for researchers who seek to efficiently extract relevant and high-quality information from vast repositories of data[1]. Deep learning methods have been applied to medicine fields, including health records, oncology etc [2], [3]. As large language models (LLMs) gain prominence in academic and industrial applications [4], [5], [6], their potential as research facilitators is undeniable. However, building reinforcement-learning-based LLMs that can effectively assist researchers requires not only advanced model architectures but also carefully curated data infrastructures. Summarizing research data in a structured and efficient manner, compatible with retrieval-augmented generation (RAG), is a fundamental step toward achieving cost-effective and scalable solutions [8].

To address this need, we propose a fine-tuning based systematic approach to develop customized QA datasets tailored to RAG-based querying mechanisms. LoRA (Low-Rank Adaptation) is a tuning technique that enables efficient fine-tuning of large language models by updating only a subset of the model's parameters in a low-rank matrix format in a cost-effective manner [7]. Fine-tuning datasets serve as the backbone of LLM-based research assistants, enabling domain-specific fine-tuning and enhanced query responses. Generating a reliable QA database requires leveraging real-world data to ensure contextual relevance and accuracy. Fine-tuning LLMs on these curated datasets is a proven methodology to enhance their performance in niche domains, making it a critical component of our framework.

In this study, we present **iTRI-QA**, a tool designed to streamline the process of creating customized QA databases for research-oriented applications. Our tool employs a four-step methodology: (1) curating high-quality QA examples, (2) collecting and structuring a comprehensive research paper database, (3) fine-tuning LLMs using domain-specific, high-quality QA examples, and (4) generating a QA database optimized for RAG-based querying. Additionally, we conducted benchmarking experiments to evaluate the performance and efficacy of our approach, demonstrating its potential to transform the way researchers interact with and utilize scientific knowledge.

This work lays the foundation for a broader suite of tools aimed at supporting RAG-based customized LLMs and reinforcement-learning-powered research assistants. By focusing on efficient data structuring and domain-specific fine-tuning, **iTRI-QA** represents a critical step toward building intelligent and cost-effective research facilitators for the scientific community.

# Result

To evaluate the performance and efficacy of **iTRI-QA**, we conducted a series of experiments focused on benchmarking the tool's capabilities in generating customized QA databases and facilitating RAG-based LLM querying. The results highlight the robustness of the four-step methodology and its potential to streamline research workflows. Below, we summarize key findings from our evaluation.

**Curate High-Quality QA Examples**

The QA example generated by **iTRI-QA** in step 1 was designed to create an accessible and efficient platform for curating QA examples, complete with metadata like PMID, DOI, and categories. The project is built using a modularized Flask application, separated into distinct components for scalability and maintainability. The front-end interface, built with HTML and styled with CSS, provides a user-friendly form for submitting QA pairs and displays real-time progress statistics, such as the total number of submissions and their distribution across categories **(Figure 2A, B)**. Data is saved in a lightweight JSONL format, making it easy to store and retrieve examples. For accessibility, the project is packaged with dynamic configurations, allowing users to customize the hosting URL and output file path via command-line arguments.

**Collect A Comprehensive Research Paper Database**

To streamline the process of managing research paper metadata, we designed this database collection utility by integrating multiple input formats into a unified and efficient storage framework. The system supports input from BibTeX (.bib) and NBIB (.nbib) files, commonly used formats for academic references, ensuring compatibility with widely adopted reference management tools. The input data is parsed and transformed into a standardized structure before being stored in highly efficient output formats like YAML files or a MongoDB database. YAML is chosen for its readability and portability, making it ideal for small-scale or configuration-centric use cases. On the other hand, MongoDB, a NoSQL database, is employed for large-scale datasets due to its scalability and robust querying capabilities, especially with an index built on DOI id for fast and precise retrieval. The framework includes modular functions to read, process, and merge data from diverse sources, with built-in options to export the processed data to various formats **(Figure 2C)**. This design not only enables easy integration into existing workflows but also ensures memory-efficient storage and quick retrieval, making it an indispensable tool for researchers managing large bibliographic datasets.

## Fine-tune LLMs using domain-specific, high-quality QA examples

To facilitate fine-tuning for a custom medical question-answering (QA) dataset, we designed a Python-based script leveraging the ITRI-QA framework with LoRA (Low-Rank Adaptation) tuning for efficient parameter updates. The script dynamically integrates model initialization, prompt rendering, and QA generation processes. LoRA tuning enables the modification of a subset of the model's parameters, significantly reducing computational and storage requirements while maintaining high performance. A PromptManager module is employed to structure and manage template-based prompts, tailored to extract contextual information from abstracts stored in JSONL format. The ITRI-QA generates responses by processing the rendered prompts, enabling the creation of domain-specific QA pairs. This modular and error-handling-aware design ensures flexibility, robustness, and ease of integration into iterative QA dataset generation workflows. By coupling efficient context parsing with scalable model operations, enhanced by LoRA tuning, the framework streamlines the adaptation of large language models for specialized medical research applications **(Figure 2D)**.

## Generate a QA database using Fine-tuned LLMs

The fine-tuned LLM is deployed to generate an initial set of QA pairs by processing domain-specific research data. The model leverages its training to extract relevant questions and answers aligned with the study's methodology, findings, and key insights. The output is a machine-generated QA database, providing a scalable and efficient method for creating structured question-and-answer datasets directly from scientific abstracts and manuscripts. To ensure the generated QA database meets high-quality standards, a post-processing phase is applied. This step involves cleaning, trimming, and formatting the QA pairs to eliminate inconsistencies and improve readability. Human experts with domain expertise evaluate the QA pairs to assess their relevance, accuracy, and clarity. This manual review process ensures that the database maintains scientific rigor and is suitable for downstream applications **(Figure 2E)**.

## Benchmarking Different Size of QA Examples in LLM Models

To profile the performance of iTRI-QA in PubMed database, we sampled QA datasets at different sizes and compared the fine-tuning performance using the evaluation loss score. The evaluation loss for the fine-tuned Llama-3.2-1B and Llama-3.2-3B models was assessed across QA sizes of 3, 5, 8, 10, and 25, revealing a consistent reduction in loss as QA size increased. While the 1B model performed slightly better at QA size 3 (13.31 vs. 13.71), the 3B model consistently outperformed the 1B model at larger QA sizes, achieving lower losses such as 12.57 vs. 13.22 at size 8 and 9.13 vs. 9.27 at size 25. These findings demonstrate the superior scalability and effectiveness of the Llama-3.2-3B model in fine-tuning tasks with larger datasets **(Figure 3A).**

We also collected human expert evaluation on the QA database generated by Llama-3.2-1B and Llama-3.2-3B. The manual evaluation of QA pairs was conducted using a scoring matrix to assess format adherence, question accuracy, answer accuracy, length, and category alignment. Total scores ranged from 0 to 15, reflecting the quality of QA pairs reviewed by human experts. For the Llama-3.2-1B model, the score range is from 2 a 9. QA pairs that failed to meet format requirements (format score = 0) sometimes achieved high accuracy scores (4 each for question and answer) but were penalized in other dimensions. Pairs adhering to the format showed more variability, with some scoring low due to inaccuracies and others achieving higher scores based on content quality. The Llama-3.2-3B model produced total scores ranging from 2 to 10, with a notable proportion of QA pairs achieving perfect scores due to high accuracy and proper adherence to format and length requirements **(Figure 3B).** While some pairs scored as low as 2 due to low content accuracy, the 3B model outperformed the 1B model overall, generating a higher proportion of high-quality QA pairs. This highlights the 3B model's superior accuracy and consistency, making it better suited for fine-tuning and deployment. However, both models exhibited variability, underscoring the importance of thorough manual evaluation to ensure the QA dataset's reliability.

## Method

**Data Curation**

Pubmed Research Literature

To ensure a robust and comprehensive dataset for this study, research papers were systematically collected from PubMed. A targeted search strategy was employed using the keywords "National Health and Nutrition Examination Survey" and "Telomere" to identify studies at the intersection of population health and telomere biology. All abstracts retrieved through this search were downloaded and stored in *.bib format, a standardized format for managing bibliographic data to ensure abstract, PMID, DOI, and other metadata variables are saved.

Customized QA Set Curation

The curation of a high-quality, customized QA dataset is achieved by loading QA toolset locally to ensure data privacy and efficient processing. Each paper in the database was reviewed by scientists with PhD-level expertise in the relevant fields. These domain experts generated QA pairs based on either the abstract or the full manuscript, depending on the complexity and depth of the content. Questions were categorized into three types to ensure diversity: knowledge-based questions for key findings and theoretical insights; methodological questions for experimental design and data collection; and discussion-based questions for implications and limitations. In total, 50 QAs were collected for this QA set.

**Framework Development**

High-quality QA Submission Webtool

The QA Submission Webtool is a modular Flask-based application designed to streamline the submission and management of QA pairs. The application is divided into distinct modules to ensure scalability and ease of maintenance. Route Definitions (utils/submit_qa_sample/routes.py) to map and manage the endpoints for the web interface. Utility Functions (utils/submit_qa_sample/utils.py) to handle file I/O, data validation, and metadata processing. Configuration Settings (utils/submit_qa_sample/config.py) for flexible adjustments to hosting parameters and output paths. Main Application (utils/submit_qa_sample/app.py) provides command-line options for deploying the webtool, including dynamic configuration of hosting URLs and output file paths.

The front-end interface is built using HTML and styled with CSS, offering an intuitive form for submitting QA pairs. Features include three parts. A real-time progress tracking part to display the total number of submissions and their distribution across predefined categories (e.g., knowledge, methodology, discussion). A data storage interface to save submissions in JSONL format to ensure efficient storage and retrieval. A command line input so users can customize deployment by specifying parameters such as host IP, port number, and output file location. Here is an example:

```
$ python utils/submit_qa_sample/app.py --host 127.0.0.1 --port 8080 --file my_data.jsonl
```

Reference Library Database Creation

The Reference Library Database Creation process is designed to integrate metadata from diverse academic sources into a unified and accessible format. Input files, including BibTeX (.bib) and NBIB (.nbib), are parsed using a Python-based script that extracts essential metadata such as titles, authors, publication dates, DOIs, and abstracts. The data is then standardized into two output formats based on the use case: YAML and MongoDB. The YAML format is prioritized for smaller datasets or scenarios requiring human readability and portability. It serves well for configuration purposes or single-session data analysis. Conversely, for large-scale datasets, MongoDB is used due to its robust querying capabilities and scalability. Each entry is indexed by DOI to enable quick and precise retrieval. The system includes automated deduplication and merging of records from multiple sources, ensuring data consistency and eliminating redundancies. To covert the bib reference library to a feedable YAML file, run:

```
$ python main.py --bibtex_files pubmed-set.bib --output_type yaml --yaml_file output.yaml
```

Prompt Engineering

To generate domain-specific QA pairs, we developed a structured prompt template designed to extract detailed and relevant information from scientific abstracts. The prompt utilizes a JSON-based structure and follows a step-by-step guideline to ensure consistency and specificity in question-answer generation. Each prompt includes 3 clear instructions; focus on extracting specific details about the study's methods or findings; avoid generic or logic-based questions to maintain relevance; output results in a JSON format with defined keys for "question" and "answer." The prompt was optimized for precision and contextual relevance by iteratively testing on a diverse range of abstracts.

Language Model Fine Tuning

To fine-tune the language models for domain-specific QA generation, we utilized the iTRI-QA framework combined fine-tuned LLM with LoRA. LoRA allows for efficient parameter updates by modifying only a small subset of the model's weights, significantly reducing computational and storage costs while maintaining high performance. The fine-tuning process involved: model initialization, prompt integration, and training configuration. Model initialization selects a pretrained base model, while prompt integration employs a PromptManager module to dynamically render templates designed to extract contextually relevant QA pairs from domain-specific abstracts. Utilizing an argparse interface, users can specify the model variant to employ, with the system defaulting to GPU-based execution if available. Training configuration utilizes GPU-based execution where available, with options for users to specify model variants and resource allocation.

**Customized QA Database Generation and Benchmarking**

Curated QA Selection and Preparation

We utilized a JSONL file containing QA data as the primary dataset. For evaluation, a separate JSONL file was employed. The training data was sampled at different sizes ('3', '5', '8', '10', and '25' QA pairs) to investigate the model's performance under varying data constraints. The raw data was structured into a format with "context," "question," and "answer" fields. The data was tokenized using the Hugging Face AutoTokenizer, with a maximum token length of 512. Prompts were generated using the PromptManager with a pre-defined template (llama3.2.j2). Each entry was processed into input-target pairs suitable for causal language modeling.

Fine-tune with LoRA Configuration

To optimize memory and computation, we implemented Low-Rank Adaptation (LoRA) using Causal Language Modeling (TaskType.CAUSAL_LM). The rank was set to 16 to enable better task-specific learning, and the alpha value was configured to 32, providing an effective scaling factor for fine-tuning. Additionally, a dropout rate of 0.1 was applied to mitigate overfitting on smaller datasets.The LoRA configuration was integrated with the base model to enable efficient parameter updates during fine-tuning.

Model Fine-tuning with LoRA

We fine-tuned two LLaMA models: meta-llama/Llama-3.2-1B and meta-llama/Llama-3.2-3B. Training was conducted with a learning rate of 3e-5 and a batch size of 1 for both training and evaluation. Gradient accumulation was set to 8 steps, and the models were trained for 5 epochs. Weight decay was configured at 0.1 to prevent overfitting. Evaluation was conducted on an epoch-based strategy, with the best model saved based on its evaluation performance. FP16 computation was enabled when GPU support was available to enhance computational efficiency. Additionally, early stopping was applied with a patience of 3 epochs to prevent overtraining. Fine-tuning was executed using the Hugging Face Trainer API. The model and tokenizer were saved after each experiment for reproducibility.

Automated Evaluation and Logging

The fine-tuned models were evaluated on the QA dataset using the Trainer's evaluation functionality after each epoch. This evaluation utilized a fixed dataset consisting of 100 augmented samples. For each evaluation run, the loss was calculated based on the evaluation dataset, and early stopping was triggered when the evaluation loss no longer showed improvement, ensuring the model avoided overfitting. For each model and data size combination, evaluation results (e.g., accuracy) and other metrics were logged to a results file. Additionally, model-specific evaluation outcomes were stored in a structured format for subsequent analysis. Experiments were performed across various training set sizes and model configurations to assess generalization capabilities and data efficiency. Temporary files for sampled datasets were created and deleted during each iteration to optimize disk usage.

Human Expert Evaluation

To evaluate the quality of the QA pairs, we conducted a manual assessment using a scoring matrix designed to measure five key dimensions: format adherence, question accuracy, answer accuracy, length, and category alignment. Human experts assigned scores to each criterion based on specific guidelines. For format adherence, a score of 0 indicated that the QA pair did not meet format requirements, while a score of 2 denoted compliance. Question and answer accuracy were scored as 0 for incorrect content, 2 for errors or missing information, and 4 for fully correct and contextually appropriate content. Length was assessed with a score of 0 for pairs that were excessively long or short and 1 for proper length. Category alignment received a score of 0 for misclassified pairs and 4 for correctly categorized ones. Each QA pair was reviewed independently by multiple experts, and scores were aggregated to identify strengths and weaknesses in the QA dataset. This thorough evaluation ensured the dataset's suitability for fine-tuning and deployment processes.

## Data and Code Availability

Data and code used in this dataset has been deposited in GitHub(https://github.com/Liu-Qiming/iTRI-QA/releases/tag/v1.0.0), analysis code is upon request to the corresponding authors.

## Discussion

The advent of large language models has transformed numerous fields by enabling advanced natural language processing and knowledge extraction capabilities. However, their application in the research domain poses unique challenges. Unlike generalized applications, the research field demands a high degree of data accuracy, regulatory compliance, and domain-specific precision. Additionally, data privacy is paramount, necessitating the development of customized LLMs and medium-sized models that can be deployed locally. These models strike a balance between computational efficiency and the ability to meet the rigorous demands of the research community.

An accessible and adaptable framework, such as iTRI-QA, is crucial to democratize the use of LLMs in research. By providing researchers with a structured pipeline to create customized QA databases, our framework facilitates the deployment of RAG-based LLMs. The generation of custom QA datasets is particularly critical, as it directly influences the relevance and utility of RAG models in answering domain-specific queries. This approach enables researchers to efficiently curate, retrieve, and synthesize knowledge from large datasets, paving the way for broader adoption of LLMs across various scientific disciplines.

Despite its potential, the framework also has limitations. The computational resources required for fine-tuning and deploying even medium-sized LLMs can be prohibitive for individual labs with limited infrastructure. Additionally, the setup and integration of such systems require technical expertise, which may present a barrier to entry for researchers without a computational background. Addressing these challenges through improved hardware access, cloud-based solutions, and user-friendly interfaces will be essential for the widespread adoption of such tools.

Future work should focus on optimizing the computational requirements of LLM frameworks while maintaining performance and accuracy. Collaboration between AI developers and domain experts can also accelerate the customization and usability of these tools, enabling researchers to derive meaningful insights while ensuring data privacy and regulatory compliance. By addressing these challenges, frameworks like iTRI can play a pivotal role in advancing the adoption of LLMs in research and facilitating data-driven innovation.

# Figure Legend

**Figure 1 Framework of iTRI-QA Design for Customized QA Database Generation.**

**Figure 2 iTRI-QA Framework and Interface**
(A) The user interface for submitting QA pairs, including fields for question, answer, PMID, DOI, and category. A real-time progress summary is displayed, showing the number of QA pairs submitted and their distribution across categories (e.g., Knowledge, Method, Discussion). (B-E) Break-down steps for each step of the iTRI-QA framework.

**Figure 3 iTRI-QA Benchmarking**
(A) The bar plot illustrates the fine-tuning evaluation loss for the Llama-3.2 1B and 3B models across varying QA sizes (3, 5, 8, 10, and 25). The evaluation loss represents the performance metric during the fine-tuning process, with lower values indicating better model performance. Each bar corresponds to a specific QA size, with blue bars representing the 1B model and orange bars representing the 3B model. (B) Box plot comparing the total human evaluation scores for QA datasets generated by two fine-tuned Llama models: Llama-3.2-1B and Llama-3.2-3B. The evaluation scores were based on criteria such as relevance, accuracy, and clarity, as assessed by domain experts. The whiskers represent the minimum and maximum scores, the boxes indicate the interquartile range, and the 'X' marks the mean score.

Figure 1

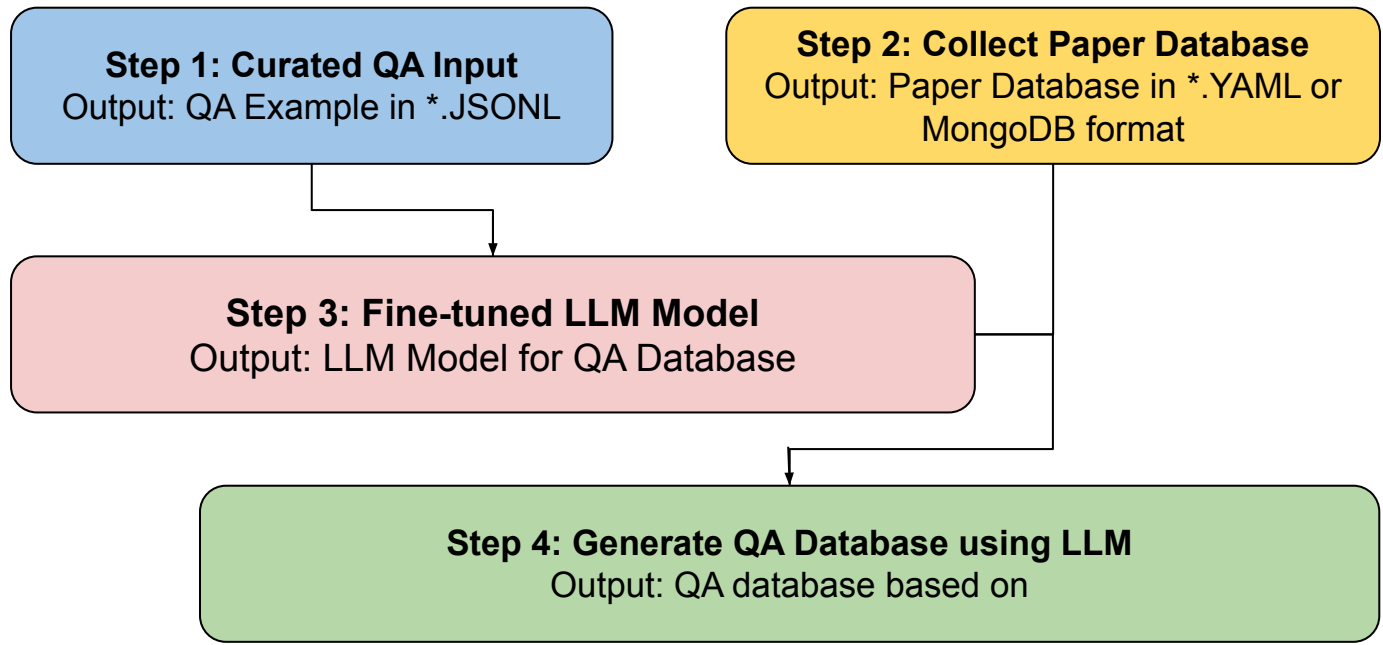

**Figure 2**

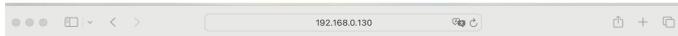

**Figure 3**

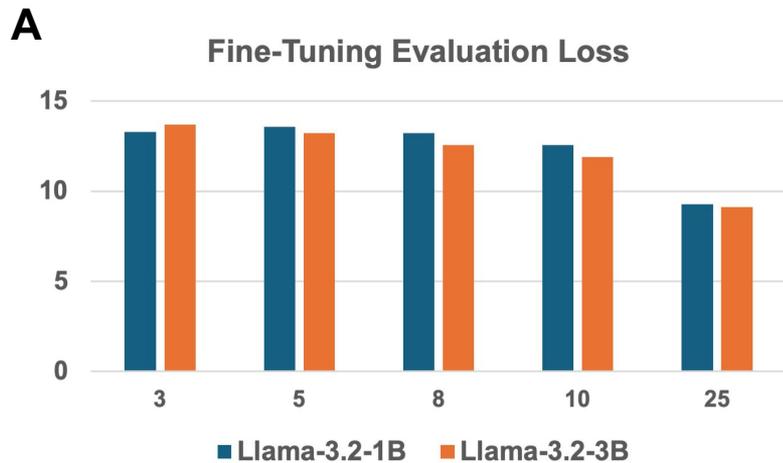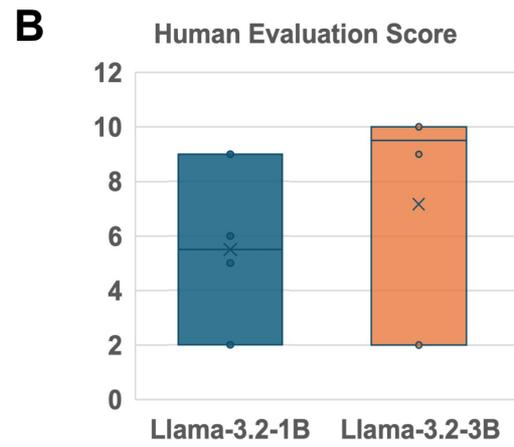